\documentclass[11pt,a4paper]{article}

\usepackage{amssymb, amsmath, amsthm}
\usepackage{graphics, color}
\usepackage{multirow}
\def\q{\hfill\rule{0.5ex}{1.3ex}}

\title{Improved approximation schemes for early work scheduling on identical parallel machines with common due date}

\author{Weidong Li \thanks{Corresponding author. E-mail: weidongmath@126.com (W. Li)}¡¡\\
School of Mathematics and Statistics, Yunnan University,
 Kunming, 650504, PR China}

\begin{document}

\maketitle

\begin{abstract}
We study the early work scheduling problem on identical parallel machines in order to maximize the total early work,
 {\it i.e.}, the parts of non-preemptive jobs executed before a common due date. By preprocessing
 and constructing an auxiliary instance which has several good properties, we propose an efficient
  polynomial time approximation scheme with running time $O(n)$,
 which improves the result in [Gy\"{o}rgyi, P., \& Kis, T. (2020). A common approximation framework for early work,
 late work, and resource leveling problems. {\it European Journal of Operational Research}, 286(1), 129-137], and  a fully
  polynomial time approximation scheme with running time $O(n)$ when the number of machines is a fixed number,
  which improves the result in [Chen, X.,  Liang, Y., Sterna, M., Wang, W., \& B{\l}a\.{z}ewicz, J. (2020b).
 Fully polynomial time approximation scheme to maximize early work on parallel machines with common due date.
 {\it  European Journal of Operational Research}, 284(1), 67-74], where $n$ is the number of jobs, and the hidden constant
 depends on the desired accuracy.

\end{abstract}
\newpage
\section{Introduction}
Early work scheduling is to schedule $n$ jobs to $m$ identical parallel machines in
non-overlapping and non-preemptive way, such that  
the total early work of jobs is maximized, where early work denotes
a part of a job executed before a common due date.  
This problem finds many practical applications in control system when collecting data from sensor,
in agriculture in the process of harvesting crops, in manufacturing systems in planning technological processes,              
and in software engineering in the process of software testing (Sterna \& Czerniachowska, 2017).

For a maximization problem, a $\rho$-approximation algorithm is  a
polynomial time algorithm which always finds a feasible solution with objective value at least
 $\rho$ times the optimal value. The supremum value of $\rho$ for which an algorithm is a
$\rho$-approximation is called the approximation ratio or the performance guarantee of the
algorithm. A polynomial time approximation scheme (PTAS, for short) for a given problem is a
family of approximation algorithms such that the family has a $(1-\epsilon)$-approximation
algorithm for any $\epsilon\in (0,1)$. An efficient polynomial time approximation scheme (EPTAS, for short)
is a PTAS whose runing time is upper bounded by the form $f(\frac{1}{\epsilon})poly(|I|)$ 
where $f$ is some computable (not necessarily polynomial) function and $poly(|I|)$ is
a polynomial of the length of the (binary) encoding of the input. A fully polynomial
time approximation scheme (FPTAS, for short) is an EPTAS which satisfies that $f$ must be a polynomial in $\frac{1}{\epsilon}$.

When $m=2$,  Sterna \& Czerniachowska (2017) proposed a PTAS based on structuring problem input.
Chen et al. (2020c)  proved that the
classical LPT (Largest Processing Time first) heuristic is an approximation algorithm with approximation ratio $9/10$. 
When $m$ is a fixed number, Chen et al. (2020b) proposed a FPTAS based on a dynamic programming approach.
When $m$ is not fixed, Gy\"{o}rgyi and Kis (2020) proposed a PTAS for a more general case.  

Chen et al. (2016) studied the online early work scheduling problem on parallel
machines, and proposed an optimal online algorithm for two identical machines. Chen et al. (2020a) also studies
 four semi-online versions of early work scheduling on two identical machines. Early work scheduling is closely related to late work minimization, 
which has been widely investigated both from theoretical aspect and
practical applications for many years (Sterna, 2011). 

In this paper,   we propose an EPTAS with running time $O(n)$ for the early work scheduling problem,
 which improves the result in   (Gy\"{o}rgyi \& Kis, 2020), and  a FPTAS with running time $O(n)$ when the number of machines is a fixed number,
  which improves the result in (Chen et al., 2020b), where $n$ is the number of jobs, and the hidden constant
 depends on the desired accuracy.
 
The remainder of this paper is organized as follows. In Section 2,
we describe the definition of the early work scheduling problem and some
preliminaries.  In Section 3, we construct an auxiliary instance and obtain some 
important properties. In Section 4, we give an EPTAS for the  early work scheduling problem. 
 In Section 5, we give an improved FPTAS for the  early work scheduling problem where the number of machines is a fixed number.
   We present some conclusions and possible directions
for future research in the last section.

\section{Preliminaries}
We are given an instance $I=({\cal J}, {\cal M}, p, d)$, which consists of a set ${\cal J}=\{J_1,J_2,\ldots, J_n\}$  of
$n$ jobs, a set ${\cal M}=\{M_1, M_2, \ldots, M_m\}$ of $m$ identical parallel machines, and a common due date $d$. Each job
$J_j$, which is described by the processing
time $p_j$, is  required to assign to one machine in
non-overlapping and non-preemptive way. The early work $X_j$  of job $J_j$ is determined by the
 job completion times $C_j$ and the common due date $d$, i.e.,
  \begin{eqnarray*}
 X_j=\min\{p_j, \max\{0, d-(C_j-p_j)\}\}.
  \end{eqnarray*}
For convenience, let $p(S)=\sum_{J_j\in S}p_j$ be the total processing time of the jobs in $S$ for any subset $S\subseteq {\cal J}$.
For $i=1,2,\ldots,m$, let $S_i$ be the set of jobs assigned to $M_i$ in a feasible solution.
 The early work scheduling problem on identical parallel machines, denoted by $P|d_j=d|X$, is to schedule all the jobs
in ${\cal J}$ to $m$ machines, in order to maximize  the total early work of all jobs
\begin{eqnarray}
 X=\sum_{j=1}^{n}X_j=\sum_{i=1}^{m}\min\{C_i, d\},
  \end{eqnarray}
where $C_i=p(S_i)$ is
the completion time of machine $M_i$ (Sterna \& Czerniachowska, 2017). If $m$ is a constant, this problem is denoted by $P_m|d_j=d|X$.

As mentioned in (Gy\"{o}rgyi \& Kis, 2020), the jobs with  processing
time $p_j\geq d$ will be scheduled on distinct machines, and can be deleted from the instance
with the machines processing them. For convenience, we assume that each job $J_j\in {\cal J}$ satisfies that
\begin{eqnarray}
\text{A1: } p_j<d.
  \end{eqnarray}

Let $OPT$ be the optimal value of a given instance $I$ for $P|d_j=d|X$. We schedule all jobs  using the classical LPT algorithm (Chen et al., 2020c) 
to obtain a feasible solution $(S_1,S_2,\ldots,S_m)$.
Let $M_{i_{max}}$ be the machine with the largest completion time. If $C_{i_{max}}\leq d$, which implies that all jobs complete
 before the common due date $d$, $(S_1,S_2,\ldots,S_m)$ is an optimal solution. If $C_{i_{max}}>d$ which implies that there are at least
 two jobs assigned to machine $M_{i_{max}}$, let $J_l$ be last job assigned to machine $M_{i_{max}}$. If $p_l>\frac{1}{2}d$, we have
$C_i\geq \frac{1}{2}d$ for any $i=1,2,\ldots,m$, as every machine is assigned at least one job with processing time no less than $p_l$ before assigning $J_l$.
If $p_l\leq \frac{1}{2}d$, we have $C_{i_{max}}-p_l> \frac{1}{2}d$, implying that $C_i\geq \frac{1}{2}d$ for any $i=1,2,\ldots,m$, as the completion time of
each machine $M_i$ is at least $C_{i_{max}}-p_l$  before assigning $J_l$, by the choice of the LPT algorithm.
Therefore, $C_i\geq \frac{1}{2}d$ for any $i=1,2,\ldots,m$ in any case, which implies that $OPT\geq \sum_{i=1}^{m}\min\{C_i, d\}\geq \frac{1}{2}md$. Thus, for  convenience, we assume that
\begin{eqnarray}
\text{A2: } \frac{1}{2}md\leq OPT\leq md,
  \end{eqnarray}
where the last inequality follows from the definition of $X$.

 If $p({\cal J})>2md$, for $i=1,2,\ldots,m-1$, we schedule
 the unassigned jobs to machine $M_i$, until the completion time $C_i$ of machine $M_i$ first exceeds
  $d$. The remaining jobs are scheduled on $M_m$. By the assumption A1, we have $d<C_i<2d$ for $i=1,2,\ldots,m-1$, and $C_m>d$, which implies that we obtain
   an optimal solution. Therefore,  for  convenience, we assume that
 \begin{eqnarray}
\text{A3: } p({\cal J})\leq 2md.
  \end{eqnarray}

\vspace{1mm}\noindent {\bf Theorem~1.}  There exists an optimal solution $(S^*_1, S^*_2, \ldots, S^*_m)$ for instance $I$ such that
 \begin{eqnarray*}
 p(S^*_i)\leq 3d, \text{ for } i=1,2,\ldots,m.
 \end{eqnarray*}
\noindent  {\bf Proof. } If there is a machine $M_{i_1}$ such that $C^*_{i_1}>3d$, by the assumptions A1 and A3, there must be a machine
 $M_{i_2}$ such that $C^*_{i_2}\leq 2d$. Without decreasing the objective value, we can reassign a job from $M_{i_1}$ to  $M_{i_2}$ until
 \begin{eqnarray*}
C^*_i\leq 3d, \text{ for any } i=1,2,\cdots,m.
  \end{eqnarray*}
Therefore, the theorem holds. \q

 \section{An auxiliary instance}
   For a given desired accuracy $\epsilon\in (0,1)$, let $\delta>0$ be a rational number such that
  \begin{eqnarray}
\delta\leq \frac{1}{10}\epsilon, \text{ and } \frac{1}{\delta} \text{ is an integer number.}
    \end{eqnarray}
 Let  ${\cal J}^{0}=\{J_j\in {\cal J}|p_j<\delta d\}$ be the set of {\it small} jobs. Furthermore, we divide the  {\it big } jobs in
 ${\cal J}\setminus {\cal J}^{0}$ into $K$ subsets ${\cal J}^{1}, {\cal J}^{2}, \ldots, {\cal J}^{K}$, where
  \begin{eqnarray}
 K=\frac{1-\delta}{\delta^2}, \text{ and } {\cal J}^{k}=\{J_j\in {\cal J}|\delta d+(k-1)\delta^2 d\leq p_j< \delta d+k\delta^2 d\}, \text{ for } k=1,2,\ldots,K.
    \end{eqnarray}
Clearly, $\cup_{k=0}^{K}{\cal J}^{k}={\cal J}$,  and ${\cal J}^{k_1}\cap {\cal J}^{k_2}=\varnothing$ for any $k_1\neq k_2$.
 For convenience, for $k=0,1,\ldots,K$, let
  \begin{eqnarray}
  |{\cal J}^{k}|=n_k,
    \end{eqnarray}
implying that $\sum_{k=0}^{K}n_k=n$.

For a given instance $I=({\cal J}, {\cal M}, p, d)$, we construct an auxiliary instance $\hat{I}=({\cal \hat{J}}, {\cal M}, \hat{p}, d)$ with
 $\hat{n}$ jobs
 and $m$ identical parallel machines  as follows, where
  \begin{eqnarray*}
{\cal \hat{J}}=\cup_{k=0}^{K}{\cal \hat{J}}^{k},  \hat{n}=|{\cal \hat{J}}|,  \text{ and }\hat{n}_k=|{\cal \hat{J}}^{k}|, \text{ for } k=0,1,\ldots,K.
 \end{eqnarray*}
The job set $\hat{\cal J}^0$ contains
\begin{eqnarray*}
\hat{n}_0=\lfloor\frac{\sum_{J_j\in {\cal J}^0}p_j}{\delta d}\rfloor
  \end{eqnarray*}
small jobs with processing time $\delta d$. For each job $J_j\in {\cal J}^k$ ($k=1,2,\ldots,K$), construct a job $\hat{J}_j\in \hat{J}^k$   such that
 \begin{eqnarray*}
\hat{p}_j=\lfloor\frac{p_j}{\delta^2 d}\rfloor \delta^2 d=\delta d+(k-1)\delta^2 d,
  \end{eqnarray*}
which implies that
 \begin{eqnarray}
p_j \geq \hat{p}_j\geq  p_j-\delta^2d\geq (1-\delta)p_j,
  \end{eqnarray}
where the last inequality follows from that $p_j\geq\delta d$ for each job $J_j\in \cup_{k=1}^{K}{\cal J}^{k}$.

\vspace{1mm}\noindent {\bf Theorem~2.}  $\hat{OPT}\geq OPT-4\delta md$, where $\hat{OPT}$ is the optimal value for instance $\hat{I}$.\\
\noindent  {\bf Proof. } Following from Theorem 1, let $(S^*_1, S^*_2, \ldots, S^*_m)$ be an optimal solution
 for instance $I$, such that $p(S^*_i)\leq 3d$ for $i=1,2,\ldots,m$. Since $p_j\geq \delta d$ for each job $J_j\in {\cal J}^k$ ($k=1,2,\ldots,K$),
we have
\begin{eqnarray}
\sum_{k=1}^K|S^*_i\cap {\cal J}^k|=|S^*_i\cap(\cup_{k=1}^K{\cal J}^k)|\leq \frac{3}{\delta}.
\end{eqnarray}
We construct a feasible solution for instance $\hat{I}$ as follows. For $i=1,2,\ldots,m$, schedule $\lfloor\frac{p(S^*_i\cap {\cal J}^0)}{\delta d}\rfloor$ jobs in  ${\cal \hat{J}}^0$ and  the jobs in $\{\hat{J}_j\in {\cal  \hat{J}}^k|J_j\in S^*_i\cap {\cal J}^k\}$ ($k=1,2,\ldots,K$) on machine $M_i$. Finally, schedule the
remaining jobs on any machine. Thus, by the construction of ${\cal \hat{J}}^k$, the completion time of machine $M_i$ is
\begin{eqnarray*}
\hat{C_i}&\geq& \lfloor\frac{p(S^*_i\cap {\cal J}^0)}{\delta d}\rfloor \delta d+ \sum_{k=1}^K\sum_{J_j\in S^*_i\cap {\cal J}^k}\hat{p}_j\\
&\geq& p(S^*_i\cap {\cal J}^0)-\delta d+ \sum_{k=1}^K\sum_{J_j\in S^*_i\cap {\cal J}^k}(p_j-\delta^2 d)\\
&=& p(S^*_i\cap {\cal J}^0)+ \sum_{k=1}^Kp(S^*_i\cap {\cal J}^k)-\delta d-\sum_{k=1}^K|S^*_i\cap {\cal J}^k|\delta^2 d\\
&\geq&  C^*_i-4\delta d,  \end{eqnarray*}
where the second inequality follows from (8) and the last inequality follows from (9). Therefore,
\begin{eqnarray*}
\hat{OPT}&\geq&\sum_{i=1}^m\min\{\hat{C_i},d\}\\
&\geq& \sum_{i=1}^m\min\{ C^*_i-4\delta d,d\}\\
&\geq&\sum_{i=1}^m\min\{ C^*_i,d\}-4\delta md\\
&=&OPT-4\delta md,
 \end{eqnarray*}
implying that the theorem holds. \q

For an optimal solution $(\hat{S}_1, \hat{S}_2, \ldots, \hat{S}_m)$ for instance $\hat{I}$, we construct a corresponding feasible solution
$(S_1, S_2, \ldots, S_m)$ for instance $I$ as follows.
 For  $i=1,2,\ldots,m$, we schedule the jobs in $\{J_j\in {\cal  J}^k|\hat{J}_j\in \hat{S}_i\cap {\cal \hat{J}}^k\}$ ($k=1,2,\ldots,K$ ) on machine $M_i$.
In addition, for $i=1,2,\ldots,m$, we schedule the jobs in  ${\cal  J}^0$ on machine $M_i$ until the
total processing time of jobs in $S_i\cap {\cal  J}^0$ first exceeds $(|\hat{S}_i\cap \hat{J}^0|-1) \delta d$. Finally, schedule the
remaining jobs on any machine. It is easy to verify that  $(S_1, S_2, \ldots, S_m)$ is a feasible solution. Moreover, we have

\vspace{1mm}\noindent {\bf Theorem~3.}  The objective value of $(S_1, S_2, \ldots, S_m)$  is at least $(1-\epsilon)OPT$.\\
\noindent  {\bf Proof. } For the feasible solution $(S_1, S_2, \ldots, S_m)$,  the completion time of
machine $M_i$ is
\begin{eqnarray*}
C_i&\geq&(|\hat{S}_i\cap \hat{\cal J}^0|-1) \delta d+\sum_{k=1}^K\sum_{\hat{J}_j\in \hat{S}_i\cap {\cal \hat{\cal J}}^k} p_j\\
&\geq& |\hat{S}_i\cap \hat{\cal J}^0|\delta d+\sum_{k=1}^K\sum_{\hat{J}_j\in \hat{S}_i\cap {\cal \hat{\cal J}}^k}\hat{p}_j- \delta d\\
&=&\hat{C}_i -\delta d,  \end{eqnarray*}
where the second inequality follows from (8).  Therefore, the objective value of  $(S_1, S_2, \ldots, S_m)$ is
\begin{eqnarray*}
OUT&=&\sum_{i=1}^m\min\{C_i,d\}\\
&\geq&\sum_{i=1}^m\min\{\hat{C_i}-\delta d,d\}\\
&\geq&\sum_{i=1}^m\min\{\hat{C_i},d\}-\delta md\\
&=&\hat{OPT}-\delta md\\
&\geq& OPT-5\delta md \\
&\geq& OPT-10\delta OPT\\
&\geq& (1-\epsilon)OPT,
 \end{eqnarray*}
where the third inequality follows from Theorem 2, the forth
 inequality follows from the assumption A2, and the last inequality follows from
 the definition of $\delta$. \q

 \section{An EPTAS for $P|d_j=d|X$}
In this section, we present an EPTAS based on an optimal algorithm for instance $\hat{I}$. 
From the construction of $\hat{\cal J}$ and the assumption A3, we have
  \begin{eqnarray}
  \hat{p}(\hat{\cal J}) \leq   p({\cal J})\leq 2md.
   \end{eqnarray}
   Similarly to Theorem 1, we have
   
   \vspace{1mm}\noindent {\bf Theorem~4.}  There exists an optimal solution $(\hat{S}_1, \hat{S}_2, \ldots, \hat{S}_m)$ for instance $\hat{I}$ such that
 \begin{eqnarray*}
\hat{p}(\hat{S}_i)\leq 3d, \text{ for } i=1,2,\ldots,m.
 \end{eqnarray*}
 
  For convenience, we introduce an auxiliary function $f: 2^{\cal \hat{J}}\mapsto R_{\geq 0}$, where
\begin{eqnarray*}
f(\hat{S})=\min \{\hat{p} (\hat{S}),d\}, \text{ for any } \hat{S}\subseteq  {\cal \hat{J}}.
  \end{eqnarray*}
Given an optimal solution $(\hat{S}_1, \hat{S}_2, \ldots, \hat{S}_m)$ for instance $\hat{I}$,
 the objective value  is
\begin{eqnarray}
\hat{OPT}= \sum_{i=1}^m \min \{\hat{C}_i,d\} =\sum_{i=1}^m\min \{\hat{p} (\hat{S}_i),d\}=\sum_{i=1}^mf(\hat{S_i}).
  \end{eqnarray}

   We define a {\it valid configuration} $C\subseteq \hat{J}$, denoted by ${\bf u}(C)=(u_{0}(C), u_{1}(C), \ldots, u_{K}(C))$ where $u_{k}(C)=|C\cap \hat{\cal J}^k|$, which satisfies that
\begin{eqnarray*}
u_{0}(C)\cdot\delta d+\sum_{k=1}^K u_{k}(C)(\delta d+(k-1)\delta^2 d)\leq 3d, 
  \end{eqnarray*}
implying that 
\begin{eqnarray}
 \sum_{k=0}^K u_{k}(C)\leq \frac{3}{\delta}.
  \end{eqnarray}

  Let ${\cal C}$ be the set of all valid configurations $C$. By the above inequality (12), we have
\begin{eqnarray}
|{\cal C}|\leq (\frac{3}{\delta}+1)^K=(\frac{3}{\delta}+1)^{\frac{1-\delta}{\delta^2}+1}=O((\frac{1}{\delta})^{\frac{1}{\delta^2}}),
  \end{eqnarray}
  which is a constant.
   For any valid configuration $C\in {\cal C}$, let $x_C$ be the numbers of machines that are assigned
$C$ in an optimal solution, i.e., $x_C=\{M_i\in {\cal M}|\hat{S}_i=C\}$.   The optimal solution for instance $\hat{I}$ can be found by solving the following integer program.
  \begin{eqnarray*}
&&\sum_{c\in {\cal C}}f(C)x_{C}\\
s.t. &&\sum_{C\in {\cal C}}u_k(C)x_{C}\leq \hat{n}_k, \text{ for } k=0,1,\ldots,K,\\
&&\sum_{C\in {\cal C}}x_{C}=m,\\
&&x_{C}\in \mathbb{Z}^{+}_{\geq 0}, \forall C\in {\cal C}.
  \end{eqnarray*}
Since  the number of variables of this integer linear program is  $|{\cal C}|=O((\frac{1}{\delta})^{\frac{1}{\delta^2}})=O(1)$ and  the number of
constraints is $K+2=O(\frac{1}{\delta^2})=O(1)$, as $\delta$ is a constant that do not depend on the input.
Therefore, the above integer linear program can be solved optimally within time $O(n)$, by using the  
 Lenstra's algorithm (Lenstra, 1983), whose running time is exponential in the dimension of the program but
polynomial in the logarithms of the coefficients, where
the hidden constant depends exponentially on $\frac{1}{\delta}$.

\vspace{1mm}\noindent {\bf Theorem~5.}  The problem $P|d_j=d|X$ possesses an EPTAS  with running time $O(n)$.\\
\noindent  {\bf Proof.} Constructing the instance $\hat{I}$ can be done within time $O(n)$. An optimal solution 
 $(\hat{S}_1, \hat{S}_2, \ldots, \hat{S}_m)$  for instance $\hat{I}$
can be found within time $O(n)$, by solving the above integer program. A corresponding feasible solution $(S_1, S_2, \ldots, S_m)$ can be 
constructed within time $O(n)$ as described in Section 3. Therefore, 
the overall running time is $O(n)$, where
the hidden constant depends exponentially on $\frac{1}{\epsilon}$.
The objective value of  $(S_1, S_2, \ldots, S_m)$ is at least
$(1-\epsilon)OPT$ following from Theorem 3. Therefore, the theorem holds. \q
 \section{A FPTAS for $P_m|d_j=d|X$}
In this section, we present a FPTAS based on an optimal algorithm for instance $\hat{I}$ when the number of machines
is a fixed number. Following from the fact the processing time of each job in $ \hat{\cal J}$ is no less than $\delta d$ and (10), we have
  \begin{eqnarray*}
\hat{n}\leq \frac{2m}{\delta}.
  \end{eqnarray*}

Noting that the processing time of each job in $ \hat{\cal J}$ is an integer multiple of $\delta^2 d$ and no less than $\delta d$, an optimal 
solution for instance $\hat{I}$ can be found by using the the dynamic programming algorithm in polynomial time. 
For completeness, we present the modified dynamic programming algorithm for $P_m|d_j=d|X$ proposed in (Chen et al., 2020) as follows.\\
{\bf Algorithm $DP_m$} (Chen et al., 2020) \\
1. Set initial condition $f(0,E_1,E_2,\ldots,E_m)=0$, for $E_i\in \{0,\delta d, \delta d+\delta^2 d,\ldots, d\}$, and $1\leq i\leq m$.\\
2. Calculate recurrence function\\
 \begin{eqnarray*}
f(j,E_1,E_2,\ldots,E_m)=\max  \left\{ \begin{split}
   & f(j-1,\max \{0, E_1-p_j\},E_2,\ldots,E_m)+\min \{p_j,E_{1}\}, \\
  & f(j-1,E_1,\max \{0, E_2-p_j\},\ldots,E_m)+\min \{p_j,E_{2}\},\\
  & \ldots \\
  & f(j-1,E_1,E_2,\ldots,\max \{0, E_m-p_j\})+\min \{p_j,E_{m}\},
\end{split}\right.\\
  \end{eqnarray*}
for $1\leq j\leq \hat{n}$, $E_i\in \{0,\delta d, \delta d+\delta^2 d,\ldots, d\}$, and $1\leq i\leq m$.\\
3. Determine the optimal total early work as $f(n,d,\ldots,d)$.

\vspace{1mm}\noindent {\bf Theorem~6.}  The problem  $P_m|d_j=d|X$  possesses a FPTAS  with running time $O(n)$.\\
\noindent  {\bf Proof.} Constructing the instance $\hat{I}$ can be done within time $O(n)$. An optimal solution  $(\hat{S}_1, \hat{S}_2, \ldots, \hat{S}_m)$  for instance $\hat{I}$
can be found by using {\bf Algorithm $DP_m$}, whose running time is 
\begin{eqnarray*}
O(\hat{n}(1+\frac{1-\delta}{\delta^2})^m)=O(\frac{2m}{\delta}(1+\frac{1-\delta}{\delta^2})^m)=O(\frac{1}{\epsilon^{2m+1}}).
\end{eqnarray*}
A corresponding feasible solution $(S_1, S_2, \ldots, S_m)$ can be constructed within time $O(n)$ as described in Section 3. Therefore,
the overall running time is $O(\frac{1}{\epsilon^{2m+1}}+n)=O(n)$, which
is polynomial in $\frac{1}{\epsilon}$.
The objective value of  $(S_1, S_2, \ldots, S_m)$ is at least
$(1-\epsilon)OPT$ following from Theorem 3. Therefore, the theorem holds.   \q
\section{Conclusion}
We present an EPTAS for $P|d_j=d|X$, and a FPTAS for $P_m|d_j=d|X$, where
all the jobs has the common due date. It is interesting to design a PTAS for the early work scheduling problem where the jobs have different
due dates, or prove that this problem does not possess a PTAS.

\section*{Acknowledgement}
The work is supported in part by the National Natural Science
Foundation of China [No. 61662088], Program for Excellent Young
Talents of Yunnan University, Training Program of National Science
Fund for Distinguished Young Scholars, Project for Innovation Team
(Cultivation) of Yunnan Province, IRTSTYN, and Key Joint
Project of the Science and Technology Department of Yunnan Province
and Yunnan University [No. 2018FY001(-014)].



\section*{References} 
   Chen, X.,  Kovalev, S.,  Liu, Y., Sterna,  M., Chalamon, I., \& B{\l}a\.{z}ewicz, J. (2020a).  Semi-online scheduling on
 two identical machines with a common due date to maximize total early work. {\it Discrete Applied Mathematics}, forthcoming.\\
   Chen, X.,  Liang, Y., Sterna, M., Wang, W., \& B{\l}a\.{z}ewicz, J. (2020b).
 Fully polynomial time approximation scheme to maximize early work on parallel machines with common due date.
 {\it  European Journal of Operational Research}, 284(1), 67-74.\\
Chen, X., Sterna, M.,  Han, X., \&  B{\l}a\.{z}ewicz, J. (2016). Scheduling on parallel identical machines
with late work criterion: Offline and online cases. {\it Journal of Scheduling}, 19, 729-736.\\
  Chen, X., Wang, W.,  Xie, P.,  Zhang, X.,  Sterna, M.,  \& B{\l}a\.{z}ewicz, J. (2020c).  
Exact and heuristic algorithms for scheduling on two identical machines with early work maximization.
{\it  Computers \& Industrial Engineering}, 144, Article No. 106449.\\
 Gy\"{o}rgyi, P., \& Kis, T. (2020). A common approximation framework for early work,
 late work, and resource leveling problems. {\it European Journal of Operational Research}, 286(1), 129-137.\\
   Lenstra Jr.(1983). Integer programming with a fixed number of variables. {\it Mathematics of Operations Research},
   8(4), 538-548.\\
  Sterna, M. (2011). A survey of scheduling problems with late work criteria. {\it Omega}, 39(2),
120-129. \\
 Sterna, M., \& Czerniachowska, K. (2017). Polynomial time approximation scheme for two parallel machines scheduling
 with a common due date to maximize early work. {\it Journal of Optimization Theory and Applications}, 174,  927-944.

\end{document}